\documentclass[fleqn,10pt]{wlscirep}
\usepackage[utf8]{inputenc}
\usepackage[T1]{fontenc}
\usepackage{graphicx}
\usepackage{amsmath}
\usepackage{xcolor}
\usepackage[stable]{footmisc}
\usepackage{hyperref}
\usepackage{booktabs}
\usepackage{float}
\usepackage{threeparttable}
\usepackage{multirow}

\title{\textbf{Predicting COVID-19 Infections Using Multi-layer Centrality Measures in Population-scale Networks}}

\date{This version: \today} 

\author[1,*]{Christine Hedde-von Westernhagen}
\author[2]{Javier Garcia-Bernardo}
\author[2]{Ayoub Bagheri}
\affil[1]{Eindhoven University of Technology, Department of Industrial Engineering \& Innovation Sciences, Eindhoven, 5612 AE, The Netherlands}
\affil[2]{Utrecht University, Department of Methodology and Statistics, Utrecht, 3584 CH, The Netherlands}

\affil[*]{c.hedde.von.westernhagen@tue.nl}

\begin{abstract} 

Understanding the spread of SARS-CoV-2 has been one of the most pressing problems of the recent past. Network models present a potent approach to studying such spreading phenomena because of their ability to represent complex social interactions. While previous studies have shown that network centrality measures are generally able to identify influential spreaders in a susceptible population, it is not yet known if they can also be used to predict infection risks. However, information about infection risks at the individual level is vital for the design of targeted interventions. Here, we use large-scale administrative data from the Netherlands to study whether centrality measures can predict the risk and timing of infections with COVID-19-like diseases. We investigate this issue leveraging the framework of multi-layer networks, which accounts for interactions taking place in different contexts, such as workplaces, households and schools. In epidemic models simulated on real-world network data from over one million individuals, we find that existing centrality measures offer good predictions of relative infection risks, and are correlated with the timing of individual infections. We however find no association between centrality measures and real SARS-CoV-2 test data, which indicates that population-scale network data alone cannot aid predictions of virus transmission.

\end{abstract}

\begin{document}

\flushbottom

\maketitle

\thispagestyle{empty}

\section{Introduction}

Since the onset of the COVID-19 pandemic in late 2019, a myriad of studies has attempted to adequately model how the virus spreads within and across populations in order to predict outbreaks and assess interventions (see \cite{kong2022} for an overview). Like for all epidemic diseases, models of COVID-19 have to take into account how members of susceptible populations are interconnected. To this end, it is important to consider that human interaction takes place in multiple contexts simultaneously. One way of accommodating this structure is the use of multi-layer network models, where each layer represents a different type of interaction \cite{kivela2014}.

The centrality of a node within the network has been show to be a good predictor of its spreading capacity, i.e., the number of adjacent nodes it infects. This has been studied for networks representing a single type of interaction \cite{bucur2020, dearruda2014}\cite[ch. 10.3]{rodrigues2019}, as well as for multi-layer networks where interactions can be of multiple types, modelled by the individual layers \cite{basaras2019}. While the identification of influential spreaders in a network allows for preventive measures to attenuate large infection events, the risk of infection for an individual cannot directly be inferred from this. However, the individual infection risk is of substantial interest for epidemic scenarios. It can facilitate government interventions specifically targeted at high-risk groups, or allow citizens to receive personalized risk assessments, and thus contributes to the mitigation of the overall epidemic scenario. In the context of SARS-CoV-2, several studies already employed multi-layer networks in modelling the spread of the virus to predict outbreak size \cite{bongiorno2022, chung2021, nande2021}, but none of them have used the framework to predict individual infections, and none have used large-scale registry data on individuals. This study aims at filling this gap by using network data from over one million individuals to answer the research question:

\bigskip
\textit{How well can multi-layer centrality measures predict the risk and timing of individual infections with epidemic diseases like COVID-19?}
\bigskip

In recent years, definitions have been developed for multi-layer versions of centralities, such as PageRank \cite{halu2013}, Eigenvenctor \cite{sola2013}, and Betweenness \cite{sole-ribalta2014}. \cite{dedomenico2015} then showed that node rankings based on the respective single-layer and multi-layer centralities can differ substantially. Differences between the measures can be attributed to the increased structural complexity of multi-layer networks, resulting in spreading dynamics that cannot be observed in single-layer networks \cite{dedomenico2016, salehi2015}. 

\begin{figure}[ht]
    \centering
    \includegraphics[scale=0.55]{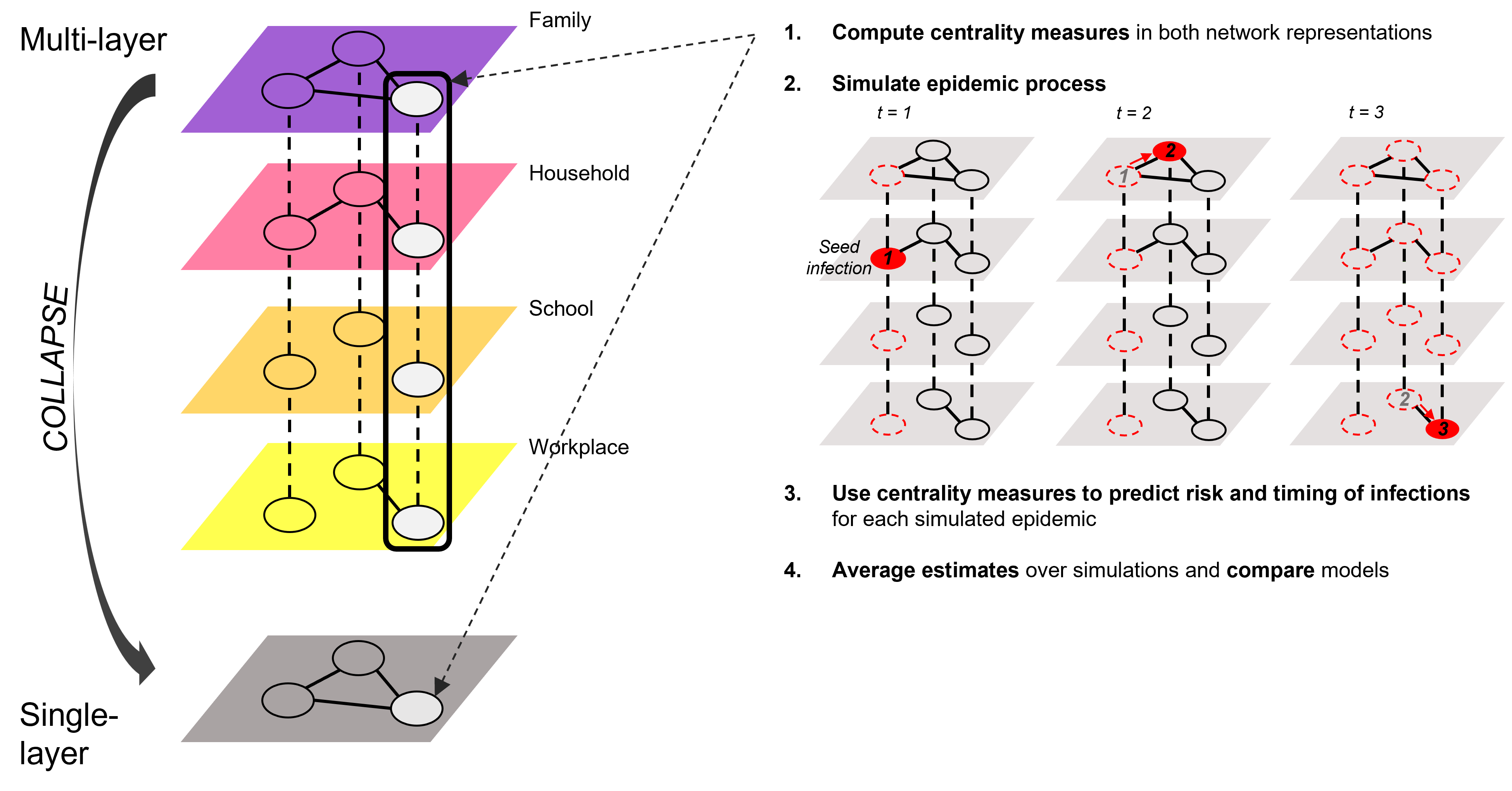}
    \caption{Schematic plot of methodological procedure based on a four-layer network of families, households, schools, and workplaces. The network contains the same three individuals as nodes that are present in each layer. Ties within layers are displayed as bold lines, ties between layers as dashed lines. Dashed inter-layer ties are categorical couplings, i.e., every node is tied to all of its representations in the other layers. A detailed description of all analysis steps shown on the right-hand side of the figure can be found in Section \ref{sec:methods}.}
    \label{fig:analysis}
\end{figure}

To assess the performance of multi-layer centrality measures in predicting infections, we made use of registry micro-data of the whole Dutch population. Administrative population networks bear a novel opportunity to researchers studying social processes, in that they do not suffer from common drawbacks of studies based on surveys or digital trace data \cite{bokanyi2023, vanderlaan2023}. Tapping this source of information thus presents another contribution of this study. Specifically, we used the registry data on family, household, school, and work relationships. On the basis of this network, we simulated the infection process of a COVID-19-like epidemic. A summary of our methodological procedure is given in Figure \ref{fig:analysis}.

The results of the prediction tasks indicate that multi-layer centrality measures can provide a good assessment of a node's relative infection risk. We also find a moderate predictive performance of certain centrality measures concerning the exact time point of infection. However, across all prediction tasks, the multi-layer measures perform slightly worse than their single-layer counterparts. Applying the measures to PCR-test data of actual COVID-19 infections showed no explanatory power, demonstrating that predicting infections at the individual level in this epidemic scenario cannot be done based on administrative network data alone.

The paper proceeds as follows: Section \ref{sec:theo} provides a theoretical background on the role of node centrality in spreading processes, and introduces concepts and notation of multi-layer networks. Section \ref{sec:data} describes the data used in this study in more detail. In Section \ref{sec:methods}, we outline the methodological procedure to answer the research question. Section \ref{sec:results} presents and discusses the results of the analyses. We close the paper with a conclusion and possible opportunities for future research in Section \ref{sec:conclusion}.

\section{Theoretical Background}\label{sec:theo}

\subsection{The role of network centrality in spreading processes}\label{sec:theo_spread}

Centrality measures are an integral part of characterizing nodes within networks, especially in the social sciences where the concept originated in the late 1940's \cite{freeman1978}. Each of the of the measures developed since then intends to capture the relevance of a node in a theoretically distinct way.

When investigating spreading processes in a network of individuals, the outcomes of interest are often aggregate results of the process, such as the final size of the spread, the rate at which it occurred, or the spreading capacity of each individual. These outcomes are indispensable in understanding the scale of spreading phenomena. Accordingly, the role of centrality measures in applications of disease, information, or behaviour spreading has mostly been to detect individuals who influence changes in those target variables \cite{bucur2020, dearruda2014, rodrigues2019, basaras2019}.

Considering single-layer network representations, correlation-based analyses from several studies \cite{bucur2020, dearruda2014, rodrigues2019, basaras2019} suggest that most types of existing centrality measures are associated with final outbreak sizes as well as the spreading caused by individual nodes. However, those findings are highly dependent on the type and size of the network, as well as the type of spreading process which was investigated. A comprehensive study conducted by \cite{dearruda2014} indicates that for epidemic spreading Degree centrality usually shows the largest correlation with final outbreak size. \cite{bucur2020} also find evidence for the influence of PageRank and Katz centrality on this outcome. That random walk based measures like PageRank perform well is further corroborated by \cite[ch. 10.3]{rodrigues2019}. For the case of multi-layer networks, research on centrality and spreading outcomes is more limited. \cite{basaras2019} demonstrated that, generally, measures based only on the direct neighborhood of a node are more predictive of its spreading capacity than measures like PageRank or Betweenness that also incorporate links at longer distances. 

While previous studies give a notion about which centrality measures are generally useful in explaining spreading processes, the spreading capacity of a node does not directly translate to its risk of becoming infected. Research on this topic is very limited, with a notable exception being \cite{christley2005}, who find quasi equal performance of Degree, Betweenness, and Closeness centrality in predicting infection probabilities. However, the study only looked at a small single-layer network, in which the centrality measures were highly correlated.

Clearly, there is no straightforward guidance on how to relate centrality with COVID-19 infections in a multi-layer network. We therefore look at three prominent measures which have shown to be important in some of the findings presented above: Degree, Eigenvector, and PageRank centrality. This leaves only measures based on calculating the shortest path between all pairs of nodes (e.g., Betweeness and Closeness) unconsidered in this paper, since it was infeasible to compute them on a network of the size presented here. Before addressing the mathematical definitions of the studied centrality measures, the concept of multi-layer networks and their notation is introduced.

\subsection{Multi-layer networks and tensorial notation}

Multi-layer networks have emerged simultaneously from a variety of research areas, each with the aim to model real-world systems composed of relationships which may differ in their type, strength, and direction. Depending on the restrictions imposed on the network, one can distinguish a broad range of subvariants of the most general multi-layer network. Examples of such restrictions are the ordering of the layers, the presence or absence of nodes in specific layers, the presence of edges between layers, or the weighting of certain edges. \cite{kivela2014} have introduced a unifying mathematical framework for multi-layer networks and provide a thorough introduction to their properties and applications. 

A concise way of notation for multi-layer networks is to represent them as tensors. Following the notation of \cite{dedomenico2015}, the \textit{adjacency tensor} of rank 4 is defined as $M^{i\alpha}_{j\beta}$, which captures the link of node $i$ in any layer $\alpha$ to node $j$ in any layer $\beta$. This formulation allows to explicitly consider multiple types of connections between nodes, which would otherwise get lost in the aggregation of the layers, but are vital to the dynamical properties of the network \cite{dedomenico2014, gomez2013}.

In this study, a \textit{multiplex network} is employed to model interpersonal relationships. This variant of a multi-layer network contains the same set of nodes in each layer, therefore it is also commonly referred to as node-aligned multi-layer network. Furthermore, it does not allow for ties between different nodes in different layers, but only between a node in one layer and its copy in all other layers, so called categorical couplings. Concerning the adjacency tensor $M$, this property can be expressed as $\alpha = \beta$, except when $i = j$. All edges in the network are undirected, representing symmetric relationships. A schematic plot of the network in the context of this study is given in Figure \ref{fig:analysis}.

\subsection{Centrality in multi-layer networks}\label{sec:centralities}

Considering the long tradition of centrality measures in network analysis, attempts have been made to establish the concept of centrality in the relatively new area of multi-layer networks. Using a tensor representation as introduced above, \cite[Supplementary Note 3]{dedomenico2015} show how some commonly used single-layer measures can be generalized to obtain analogous multi-layer measures.

In the following, we provide the equations for the multi-layer centrality measures investigated in this study. Throughout, we follow the notation of \cite{dedomenico2015} with slight deviations. For more details on the mathematical formulation of multi-layer networks and their properties see also \cite{dedomenico2013}.

\textbf{Degree Centrality}. The most commonly inspected property of a node is its degree, i.e., the number of incoming or outgoing ties of a node. Using Einstein notation for conciseness, the multi-layer degree centrality (or 'multidegree') of a node $i$ across all layers can be defined as

\begin{center}
    \begin{equation}
        \kappa_{i} = K_{i\alpha} u^{\alpha},
    \label{eq:deg}
    \end{equation}
\end{center}

where $K_{i\alpha}$ is the degree of node $i$ in layer $\alpha$ and $u^{\alpha}$ is a vector of ones used to contract the layer index, i.e., to sum up the degree over all layers. In a multiplex network as used here, this measure is analogous to the strength of a collapsed single-layer network because of the exclusively categorical couplings between layers.

\textbf{Eigenvector Centrality}. Another prominent measure of node centrality quantifies a node's importance based on the node's ties and the number of its neighbours ties. In the domain of multi-layer networks this is achieved by finding the eigentensor $\Theta_{i\alpha}$ that satisfies 

\begin{center}
    \begin{equation}
         M^{i\alpha}_{j\beta} \Theta_{i\alpha} = \lambda_1 \Theta_{j\beta},
    \end{equation}
\end{center}

where $\lambda_1$ is the leading eigenvalue of $M^{i\alpha}_{j\beta}$. Solving this eigenvalue problem and contracting the layers as in the case of Degree centrality gives the overall centrality for each node:

\begin{center}
    \begin{equation}
        \theta_{i} = \Theta_{i\alpha} u^{\alpha}.
    \end{equation}
\end{center}

\textbf{PageRank Centrality}. PageRank centrality, originally developed as part of Google's search algorithm \cite{page2001}, is based on a random walk on the network with the transition tensor

\begin{center}
    \begin{equation}
        R^{i\alpha}_{j\beta} = r T^{i\alpha}_{j\beta} + \frac{(1-r)}{NL} u^{i\alpha}_{j\beta},
    \end{equation}
\end{center}

where $r$ is the rate of jumping to a neighboring node and $1-r$ is the rate of the walker being teleported to any other node in the network. The transition tensor $T^{i\alpha}_{j\beta} = M^{k\gamma}_{j\beta} \Tilde{D}^{i\alpha}_{k\gamma}$, with $ \Tilde{D}^{i\alpha}_{j\beta}$ being the inverse of the non-zero entries of the strength tensor of the network, meaning, $ \Tilde{D}$ normalizes the adjacency tensor $M$. The number of nodes and layers are given by $N$ and  $L$, respectively. $u^{i\alpha}_{j\beta}$ is a rank-4 tensor of ones. Finally, the PageRank centrality of node $i$ is given by

\begin{center}
    \begin{equation}
        \omega_i = \Omega_{i\alpha} u^{\alpha},
    \end{equation}
\end{center}

with $\Omega_{i\alpha}$ being the eigentensor of the transition tensor $R^{i\alpha}_{j\beta}$.

\section{Data}\label{sec:data}

\subsection{Construction of the network dataset}

National registry data provided by \textit{Statistics Netherlands} (CBS) was used to construct a multi-layer network based on the Dutch population. The data provides direct information on different kinds of relationships between individuals registered as residents in the Netherlands. This allowed to assemble an undirected multiplex network that consists of four layers, namely, families, households, schools, and workplaces. Each node within a layer represents a person, and per definition of the multiplex structure, each layer contains the same set of nodes.

Unlike other approaches to obtain social network datasets, such as surveys or digital trace data, this source does not suffer from the common problems of non-response bias, selection bias, or social desirability effects. However, it should be pointed out that ties in this network do not measure social interactions directly, but only represent formal relations between individuals. While this relationship definition does not explicitly consider informal ties such as friendships, prior research has shown that most of a person's close relationships originate from, and change with, the contexts of family, school, or work \cite{buijs2022, wrzus2013}. Ties in our data thus plausibly indicate ``a highly increased probability that two individuals interact socially'' \cite[146]{vanderlaan2023}.

While country-level population data bears a highly valuable resource, it also drives up computational demands. We therefore conducted analyses on a regional subset, including all primary schools students from the Dutch capital city Amsterdam. By the beginning of 2020, the municipality (Dutch: gemeente) of Amsterdam  has housed around 5 percent of the Dutch population \cite{cbs2023}. Initiating the network construction at the school level was motivated by this context's role as bridge between different households in potential disease transmission \cite{vaniersel2023}. While the same argument could be made for work relationships, some workplaces in this dataset encompass more than a hundred employees, so the contact probability of work colleagues is on average lower than between members of a school year.

All students within the selected schools became nodes in the final network, where members of the same school year are completely interconnected. For the family relations of the network, we added all parents and full siblings of the school students as nodes. The household layer added nodes living in the same household as the students, regardless of their family relationship. We then retrieved information on the colleagues of all nodes included so far, and added them as new nodes to the network. Ties were established between people working at the same workplace address.

Data on individual COVID-19 infections based on PCR-test results of SARS-CoV-2 was also provided by CBS, and could be linked to the network data. The coverage of the SARS-CoV-2 data ranges from June 2020 until September 2021, which led to the decision to also base the network dataset on administrative records from 2020. 

\subsection{Network characteristics}

After assembling the dataset from the registry data, we arrived at a network consisting of about 1.6M nodes, with ca. 200K ties in the family layer, 273K ties in the household layer, 1.4M in the school layer, and 58.9M ties in the workplace layer. Table \ref{tab:descriptives} summarizes further network characteristics.

\begin{table}[ht]
\small
\centering
\begin{threeparttable}
\caption{Descriptive network measures by layer and for the aggregate network.}
\label{tab:descriptives}
\addtolength{\tabcolsep}{-2pt}
\begin{tabular}{@{}lllllllllll@{}}
\toprule
                &           &            &            &            &                      & \multicolumn{2}{l}{Degree Distribution} &&&\\
                \cmidrule(lr){7-11}
Layer           & Nodes     & Ties       & Clustering & Components & \% giant comp.       & Pctl. 5   & Mean     & Median & Pctl. 95   & SD \\ \hline
Family          & 166,393   & 199,781    & 0.76       & 41235      & 0.01                 & 1         & 2.40     & 2      & 4          & 1.07      \\
Household       & 164,013   & 273,344    & 1          & 41627      & 0.05                 & 1         & 3.33     & 3      & 5          & 2.07      \\
School          & 58,079    & 1,386,609  & 1          & 1566       & 0.27                 & 16        & 47.75    & 45     & 91         & 24.08     \\
Work            & 1,446,817 & 58,860,777 & 0.70       & 13022      & 0.93                 & 3         & 81.37    & 100    & 164        & 54.90     \\ \hline
Aggregate       & 1,570,812 & 60,538,563 & 0.70       & 1          & 100                  & 3         & 77.08    & 81     & 163        & 55.43     \\ \hline
\end{tabular}
    \begin{tablenotes}
      \small
      \item \textit{Note}: Instead of minima and maxima, the lower and upper five percentiles are given  for the degree distribution as to not disclose information at the individual level. 
    \end{tablenotes}
  \end{threeparttable}
\end{table}

Comparing the network sizes across layers, one can clearly observe the dominance of work relationships, which is also present in the complete population data, and results from people being employed at large companies. In assembling the original population dataset, the number co-worker ties in such companies was limited to a maximum of 100 which were sampled based on living in the same (or a close-by) location \cite[7-8]{vanderlaan2022}. This procedure resulted partly in directed relationships, i.e., $i$ being the co-worker of $j$ but not the other way around. We reconstructed these reciprocities to obtain an undirected network, which lead to some nodes having a degree $> 100$. 

Accordingly, the Degree distribution of the work layer shows large variation and is rather left-skewed. The distributions in the household, family, and school layers are much less skewed than in the work layer, but the school layer also exhibits a high amount of variation. The Degree distribution of the aggregate network is shown in Supplementary Figure S2 and strongly resembles that of the complete population as shown in \cite[Figure 1]{vanderlaan2023}.

The global clustering coefficient of $1$ in the household and school layers is caused by the design that all members of one household or school year, respectively, are completely connected. For the other layers and the aggregate network, there is also a considerable degree of clustering present, meaning, in about 70 percent of the cases where two people are linked to a third person, these two people are also connected. The clustering within families is not perfect since half-sibling relationships are not included, allowing two sampled students who share a parent to not be connected themselves. The sampling method for the work layer discussed above also leads to imperfect clustering in that layer.

While, by design of the dataset construction, the aggregate network consists of one giant component, the individual layers are fragmented into many individual components. These components represent individual families, households, school years, and workplaces. However, there is some bridging between family components present due to the mentioned half-sibling relationships. The share of nodes belonging to the giant component within a layer ranges from $0.01$ percent in the family layer to $0.93$ percent in the workplace layer.

Overall, the individual network layers and the aggregate network show quite distinct characteristics. This observation has also received a detailed discussion with respect to the complete population data in \cite{bokanyi2023}. In the context of this study, the layer heterogeneity highlights how aggregation can discard a lot of structural variation that may be vital to spreading behaviour.

\section{Methods}\label{sec:methods}

\subsection{Epidemic modeling}\label{sec:epimodel}

In order to obtain the outcome of interest for this study---individual infections---we first simulated a COVID-19-like epidemic on the network data. As commonly employed in epidemiology, we used a SIR model (see, e.g., \cite[Section 3]{salehi2015}), which captures the infection states in a population over time across three groups: \textbf{S}usceptible, \textbf{I}nfected, and \textbf{R}ecovered, where recovered individuals are permanently removed from the susceptible population. The model can be expressed as a set of differential equations:

\begin{center}
    \begin{align}
        \frac{dS}{dt} &= -\beta I S,\\
        \frac{dI}{dt} &= \beta I S - \gamma I,\\
        \frac{dR}{dt} &= \gamma I,
    \end{align}
\end{center}

where $\beta$ indicates the infection rate, i.e., the transition rate $S \rightarrow I$, and $\gamma$ the recovery rate, i.e., the transition rate $I \rightarrow R$. The state variables $S,I,R$ define the total number of individuals in the respective state, together summing up to the population size $N$. 

While, traditionally, the model assumes a homogeneous mixing of the population, when applied on a network, it is evaluated at each individual node based on the current state of its neighbors. This represents a more realistic scenario than the possibility of getting in contact with any individual of the population at any time. Instead of a fixed infection rate $\beta$, the multi-layer framework furthermore allowed that each layer got assigned its own infection rate $\tau_l$, with $l \in \{family, household, school, work\}$. These infection rates were then applied in generating new infections at weekly time steps $t$.

The values for $\tau_l$ were based on the so called secondary attack rate. This rate expresses the share of infections originating from an initial infection among the number of possibly infected individuals in a group over fixed a period of time, usually one to two weeks. Based on the literature on secondary attack rates of SARS-CoV-2 in household contexts, $\tau_{household}$ was set to $0.20$ \cite{lei2020, madewell2022, telle2021}. This means that within one week after one household member got infected, 20 percent of the household members would be infected. The same studies suggest that infections originating from children are less common, and $\tau_{school}$ was then set to $0.10$. Since family relations in this network do not necessarily imply living in the same household, $\tau_{family} = 0.15$ was set somewhat lower than $\tau_{household}$. Finally, infections in workplaces were assumed to be on average the least likely, expressed by $\tau_{work} = 0.05$.

To arrive at a vector of individual infection probabilities for the nodes in a layer $l$ in week $t$ based on the previously infected direct neighbors, we used the Reed-Frost (or chain-binomial) model \cite{abbey1952, draief2006}:

\begin{center}
    \begin{equation}
        \pmb{\tau'}_{lt} = 1 - (1 - \tau_l)^{\pmb{\Gamma}_{t-1}},
    \end{equation}
\end{center}

where $\pmb{\Gamma_{t-1}}$ is a vector with the number of infected neighbors of each node. Note that if a node became infected in one layer, it also changed its infection status in all other layers. This also implies that infection probability was higher for individuals with ties in multiple layers.

The recovery time $\gamma$ did not differ by layer and was sampled for each individual from a Weibull distribution with shape $\eta = 1$ and scale $\lambda = 5$ on a scale of days \cite{dekker2023}. Since the model was evaluated at weekly intervals, the recovery status $R_{it} \in \{0,1\}$ of a node was changed to $R_{it} = 1$ if $\sum_{t=1}^{T} I_{it-1} \times 7 > \gamma$, i.e., if the number of days infected was greater than the recovery time. In the following analyses, the model was run $k = 100$ times with differing random seed nodes from which the spread started. We set the number of seed nodes to 10 per simulation, since it is realistic to assume that the virus entered the Dutch population via more than a single infected person. This also decreased the dependence of the epidemic process on the position of the seed nodes in the network.

It should be noted that the model was constructed for a setting in the early phases of the COVID-19 pandemic, without the  possibility of vaccination, or considering differing infectiousness of virus subvariants. It was also outside the scope of this study to account for the effects of interventions like mobility restrictions. Descriptive results of the simulations are given in Supplementary Table S1 and Supplementary Figure S1.

\subsection{Prediction of infections using centrality measures}\label{sec:eval}

Before relating the centrality measures to the timing of infections, we inspected the correlations between the different measures. This enabled a first insight into qualitative differences between the measures and allowed to identify potential problems of multicollinearity in the subsequent analyses. Since the distributions of the measures did not follow normality, we used the non-parametric Spearman's $\rho$ of the centrality ranks instead of Pearson moment correlations \cite{shao2018}. We also show univariate distributions of all centrality measures in Supplementary Figures S2 through S4.

In a next step, we predicted individual infection \textit{risk} over time based on a person's centrality. For this task, we used Cox proportional hazards models, regressing time dependent infection risks on different combinations of centrality measures. The model is commonly expressed as

\begin{center}
    \begin{equation}
        h_i(t|x_i) = h_0(t) e^{\beta_1 x_{i1} + \beta_2 x_{i2} + ...},
    \end{equation}
\end{center}

where $h_i(t|x_i)$ is the so called hazard rate of an individual conditional on its values for the predictor variables $x_{i1}, x_{i2}, ...$. This can be interpreted as the instantaneous risk of an individual experiencing an event---here: infection---given that they did not experience it up until this point. The baseline hazard $h_0(t)$ is the hazard function for a person whose predictor variables are all zero. The coefficients $\beta_1, \beta_2, ...$ were estimated using the partial maximum likelihood method \cite{cox1975}, which takes into account the number of right-censored observations---i.e., individuals for which no event could be observed in the given time.

The models included the introduced centrality measures as linear combinations up to the third-order polynomial as predictors. Across-variable interactions were also considered, resulting in 25 models in total. The exact predictor combinations are displayed together with the results in Figure \ref{fig:cox-results}. The models were run separately for single- and multi-layer versions of the centrality measures to allow for comparisons between those. We evaluated the performance of the models using a variant of the Concordance index (C-index), which is weighted to account for censored observations as proposed by \cite{uno2011}. The C-index expresses the degree to which the estimated individual risks align with the order in which infections actually occurred. The index ranges from $0$, meaning the risks are completely inverse to the order of infections, to $1$, indicating perfect alignment of infection risk and order of infections.We retrieved C-index values for each of the $k$ simulations, which were then averaged after applying a Fisher \textit{z}-transformation \cite{schafer1999}. This transformation was desirable since sampling distributions of correlation-like measures do not follow normality. The transformation rescales the estimates such that they approximate a normal distribution, which then allows to construct appropriate confidence intervals.

After investigating infection risks, we obtained Spearman's rank correlations $\rho$ of the \textit{time point} of infection of a node and its centrality value for the respective multi-layer and single-layer measures. This analysis was analogous to the procedure of most studies investigating the relationship of centrality measures and outbreak size or spreading capacity. The correlation coefficient for one simulation $k$ of the epidemic was defined as

\begin{center}
    \begin{equation}
        \rho_k = \frac{cov(R(TTI_k), R(Cent))}{\sigma_{R(TTI_k)} \sigma_{R(Cent)}},
    \end{equation}
\end{center}

with $cov(R(TTI_k), R(Cent))$ being the covariance of the \textit{ranks} of the time to infection and the respective centrality, and $\sigma$ being their standard deviations. To arrive at an overall estimate of $\rho$, we again retrieved correlation coefficients for each of the $k$ simulations, which were then averaged after being rescaled to normality as mentioned for the C-index.

Lastly, we used XGBoost (Extreme gradient boosting) models to predict the time point of individual infections based on the centrality measures. XGBoost is a tree-based method which enables memory efficient identification of complex data patterns, and has been successfully employed in various machine-learning tasks \cite{chen2016}. This modeling approach estimates the actual time until an individual gets infected, as opposed to the infection risk at any point in time, as achieved by the Cox models. It also goes beyond the introduced rank correlations in that the dependent variable captures quantifiable time-intervals between infections instead of just their ordering, and by considering multiple predictors simultaneously. Since the algorithm computes estimates based on ensembles of many different decision trees, it comes at the drawback of lower interpretability than, e.g., parametric regression methods.

Within each of the $k=100$ datasets resulting from the epidemic simulations, the algorithm was trained on 10 percent of observations and then tested on the remaining 90 percent. Hyperparameters were selected using 10-fold cross-validation within one of the datasets. Final hyperparameter settings are shown in Supplementary Table S2. Model performance was assessed using $R^2$ and root-mean-square error (RMSE) values, averaged over the test partitions of the 100 simulated datasets. The importance of individual predictors was evaluated using Cover, Frequency, and Gain values, again, averaged over all simulated datasets. Gain expresses the reduction in prediction error achieved by splitting the data on a certain variable relative to the other variables, summed up over all trees. Frequency can be regarded as a more crude measure of Gain, and Cover captures the relative number of observations that were assigned to a leaf after splitting on a certain variable as compared to the other variables.

Apart from using the XGBoost algorithm with the simulated infection data, we also applied it on infections derived from positive PCR-test results of our sample. Additionally, we augmented these models with information about age, postcode area (first two digits), and whether a person is of Dutch origin, to see how much these variables could improve predictions compared to models only relying on centrality measures. Again, model performance was evaluated using $R^2$ and RMSE values. Descriptive statistics of the personal characteristics and PCR-test data are presented in Supplementary Table S4 and Supplementary Figures S6 and S7.

\subsection{Ethical statement}

Ethical approval for the use of the data was obtained by the Ethical Review Board of the Faculty of Social and Behavioural Sciences of Utrecht University on November 8, 2022 (case numbers 22-1886, 22-1887, 22-1888). Informed consent of the participants has been obtained by the parties responsible for collection and provision of the data, which are Statistics Netherlands (CBS) and the National Institute for Public Health and the Environment (RIVM). In accordance with the Statistics Netherlands Act ("Wet op het Centraal bureau voor de statistiek") and the General Data Protection Regulation of the European Union, the data is protected by strict privacy regulations ensuring that no information about individual persons is disclosed. All methods were carried out in accordance with relevant guidelines and regulations.

\section{Results}\label{sec:results}

\subsection{Correlations between centrality measures}\label{sec:correlations}

The results of the preliminary correlation analysis revealed interesting insights into the association between both multi- and single-layer centralities (Table \ref{tab:cor_within}). Across both network types, a similar pattern emerged: PageRank centrality and Eigenvector centrality showed the smallest correlation ($\rho_{multi} = 0.20, \rho_{single} = 0.33$), followed by medium-sized correlations of Degree and Eigenvector centrality ($\rho_{multi} = 0.47, \rho_{single} = 0.42$). This is likely due to Eigenvector centrality allocating most of the centrality to few highly connected nodes, resulting in little variation within the measure. The correlation of Degree and PageRank was in both cases substantial, with $\rho_{multi} = 0.79$ and  $\rho_{single} = 0.84$. 

\begin{table}[ht]
\centering
\begin{threeparttable}
\caption{Rank correlations between different centrality measures measured by Spearman's $\rho$.}
\label{tab:cor_within}
\begin{tabular}{ll|llllll}
\hline
                                                   &             & \multicolumn{3}{l|}{Multi-layer}  & \multicolumn{3}{l}{Single-layer} \\ \cline{3-8} 
                                                   &             & Degree & Eigenvector & PageRank   & Degree  & Eigenvector & PageRank \\ \hline
\multicolumn{1}{l|}{\multirow{3}{*}{Multi-layer}}  & Degree      & 1.00   &             &            &         &             &          \\
\multicolumn{1}{l|}{}                              & Eigenvector & 0.47   & 1.00        &            &         &             &          \\
\multicolumn{1}{l|}{}                              & PageRank    & 0.79   & 0.20        & 1.00       &         &             &          \\ \cline{1-1}
\multicolumn{1}{l|}{\multirow{3}{*}{Single-layer}} & Degree      & 1.00   & 0.47        & 0.79       & 1.00    &             &          \\
\multicolumn{1}{l|}{}                              & Eigenvector & 0.42   & 0.64        & 0.29       & 0.42    & 1.00        &          \\
\multicolumn{1}{l|}{}                              & PageRank    & 0.84   & 0.25        & 0.96       & 0.84    & 0.33        & 1.00     \\ \cline{2-8}
\hline
\end{tabular}
\end{threeparttable}
\end{table}
 
Looking at the correlations between the two centrality types (lower left of Table \ref{tab:cor_within}), revealed that the respective centrality counterparts of PageRank are very similar to one another, with $\rho = 0.96$. Eigenvector was also notably correlated at $\rho = 0.64$. Since Degree centrality in the multi-layer definition corresponds to the weighted Degree in the single-layer representation, those measures were perfectly correlated. Similarly, correlations of Degree centrality with any of the other measures did not differ depending on using the multi- or single-layer representation.

These results suggest that there might not be a great benefit in using the multi-layer versions of centrality, since they seem to measure similar properties as the single-layer versions. The following multivariate analyses confirm this suspicion.

\subsection{Centrality measures correlate strongly with the \textit{risk} of infection}

\begin{figure}[ht]
    \centering
    \includegraphics[width=\textwidth, trim=.08cm 1cm .05cm .05cm, clip]{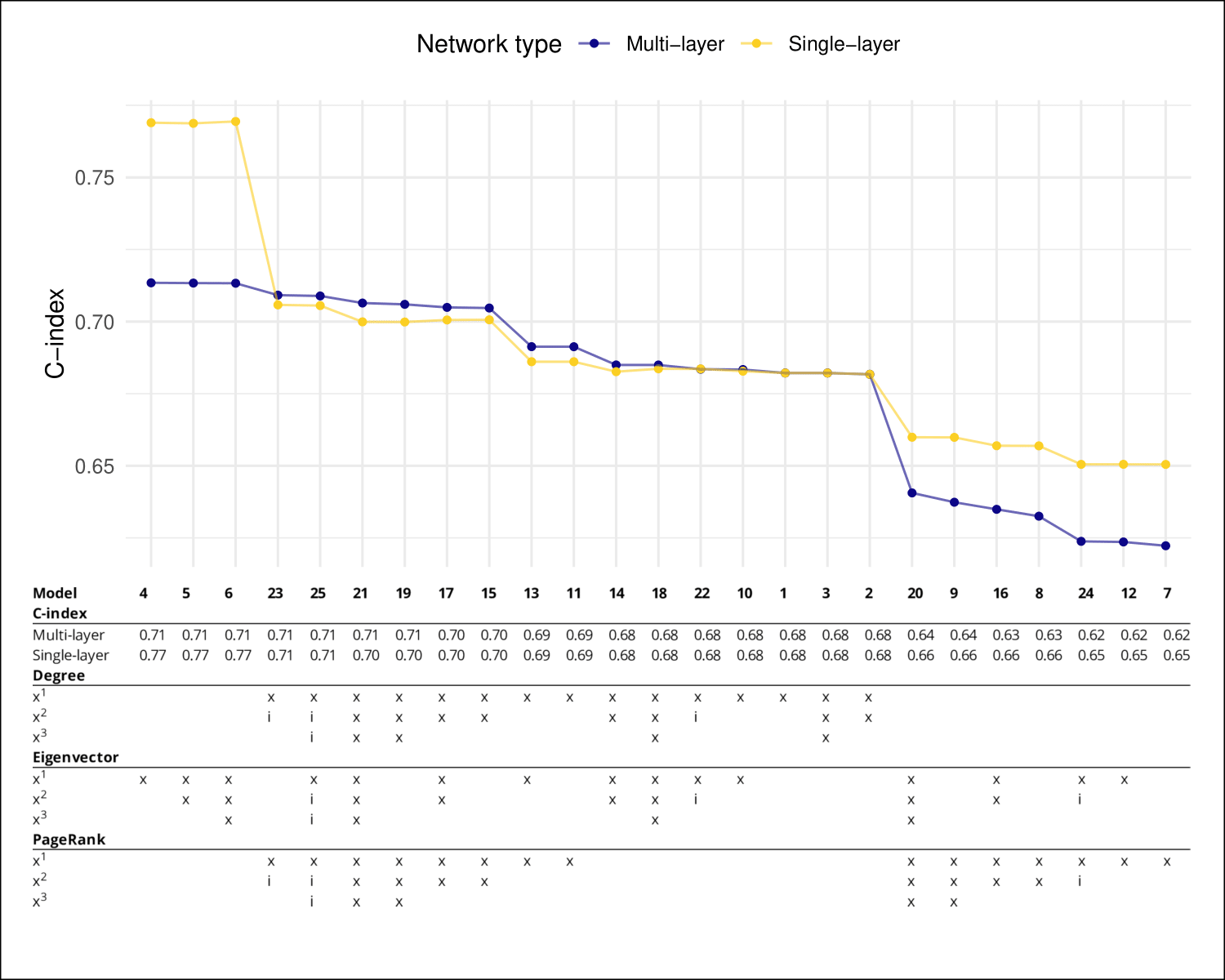}
    \caption{Concordance indices for Cox proportional hazards models by type and order of included centrality measure. An \textit{x} denotes terms composed of a single variable, \textit{i} denotes two- and three-way interactions between multiple variables. Since the estimates' confidence intervals resulting from the simulations were very narrow, they were omitted from this figure, and included in Supplementary Table S2 instead.}
    \label{fig:cox-results}
\end{figure}

We used Cox proportional-hazards models (Section \ref{sec:eval}) to examine whether centrality measures can predict infection risks over time, taking into account possible multivariate associations between the measures. C-index values of the models to assess their predictive performance are presented in Figure \ref{fig:cox-results}.  Since confidence intervals were extremely narrow across all models, they are included only in Supplementary Table S2. This shows that there was little dependence of the infection process on the seed nodes of the epidemic.

We find a strong association of up to $C=0.77$ between the centrality of the node and its relative risk of infection. This means, when choosing two random individuals in the network, the models could predict correctly who gets infected first in the epidemic in 77 percent of the cases. The risk of infection is mostly dependent on Eigenvector centrality. Even the simplest model including only linear Eigenvector centrality (Model 4)  achieved a C-index of $0.77$. 

Adding higher-order terms of any centrality did not lead to notable improvements of predictions for both multi-layer and single-layer models. Models including only PageRank centrality, or additionally also Eigenvector centrality, fared worst, with a predictive performance ranging from $C = 0.62$ to $C = 0.66$. While the ordering of centrality measures in terms of performance is consistent across multi- and single-layer models---Eigenvector $>$ Degree $>$ PageRank---the single-layer measures generally performed slightly better than the multi-layer ones.

While the good performance in risk predictions across most of the measures is a promising result, these risks only express the relative ordering of individuals in getting infected and might be of limited use in practice. In contrast, the following results uncover the extent to which the timing of the infections could be predicted accurately.

\subsection{Centrality measures correlate moderately with \textit{time} to infection}

\subsubsection{Bivariate rank correlations}

Turning from infection risks to their timing, we first inspected the rank correlation between centrality measures and the time to infection in the epidemic model. All types of centralities showed a negative Spearman's $\rho$ (Table \ref{tab:cor_between})---i.e., nodes with higher centralities become infected earlier. For both multi- and single-layer measures, Eigenvector centrality had the largest association, with $\rho_{multi} = -0.45$ and $\rho_{single} = -0.58$, respectively. Degree centrality came second with $\rho_{multi} = \rho_{single} = -0.24$. PageRank centrality followed closely after, with $\rho_{multi} = -0.15$ and $\rho_{single} = -0.19$.

\begin{table}[ht] 
\small
\centering
\begin{threeparttable}
\caption{Spearman's $\rho$ of centrality measures and time to infection averaged across epidemic simulations.}
\label{tab:cor_between}
\begin{tabular}{ll|ll}
\hline
                                                   &             & Mean  & 95\% CI         \\ \hline
\multicolumn{1}{l|}{\multirow{3}{*}{Multi-layer}}  & Degree      & -0.24 & [-0.242,-0.239] \\
\multicolumn{1}{l|}{}                              & Eigenvector & -0.45 & [-0.455,-0.448] \\
\multicolumn{1}{l|}{}                              & PageRank    & -0.15 & [-0.156,-0.154] \\\cline{1-1}
\multicolumn{1}{l|}{\multirow{3}{*}{Single-layer}} & Degree      & -0.24 & [-0.242,-0.239] \\
\multicolumn{1}{l|}{}                              & Eigenvector & -0.58 & [-0.585,-0.576] \\
\multicolumn{1}{l|}{}                              & PageRank    & -0.19 & [-0.187,-0.185] \\
\hline
\end{tabular}
\end{threeparttable}
\end{table}

These results show no support for multi-layer measures being better predictors of infection timing than their single-layer versions---but rather the opposite. Furthermore, while a relationship between the respective centralities and the time to infection could be detected in both network structures, the correlations are far lower than what has previously been observed for outbreak size or spreading capacity of a node, where $\rho$ often ranges between $0.7 - 0.9$ (see references Section \ref{sec:theo_spread}). This is an important finding, as it could indicate that the goals of possible interventions---preventing spread or preventing infections---cannot necessarily be achieved by the same means. It also hints at how the type and size of network considered in investigating spreading outcomes may lead to substantially different insights about node properties like centrality.

\subsubsection{XGBoost models with simulated infection data}

\begin{figure}[ht]
    \centering
    \includegraphics[width=\textwidth, trim=1cm 0.5cm 1cm 0cm, clip]{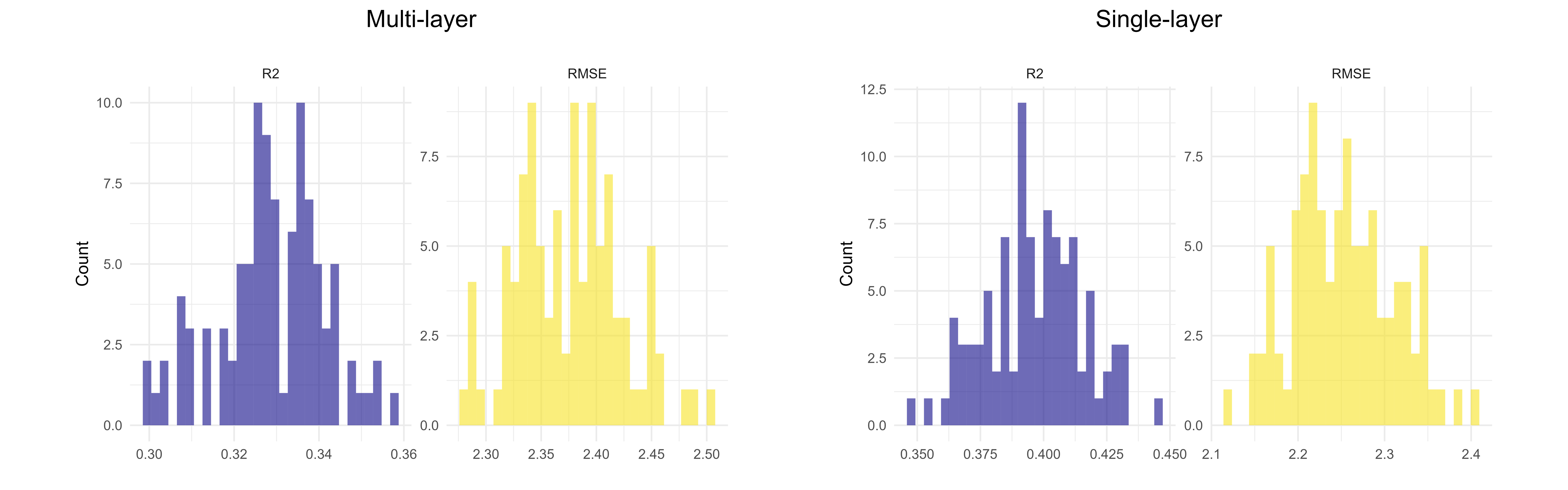}
    \caption{Distributions of $R^2$ and RMSE values across 100 epidemic simulations for XGBoost models including multi-layer or single-layer centrality measures.}
    \label{fig:xgb-metrics}
\end{figure}

The highly flexible XGBoost algorithm allows to model complex patterns underlying the relationship of centrality measures and time to infection in our data. The distributions of $R^2$ and RMSE values from the 100 simulations for models including multi-layer or single-layer centrality measures are displayed in Figure \ref{fig:xgb-metrics}.

The plot shows that the single-layer centrality measures achieved a slightly better performance in predicting the timing of infections. On average, the single-layer models performed at $R^2 = 0.40, RMSE = 2.25$, compared to $R^2 = 0.33, RMSE = 2.38$ when using the multi-layer centralities. Both centrality types exhibit similarly normally-shaped distributions of their performance metrics, showing that the result are robust to changes in the epidemic seed.

\begin{figure}[ht]
    \centering
    \includegraphics[scale=.9]{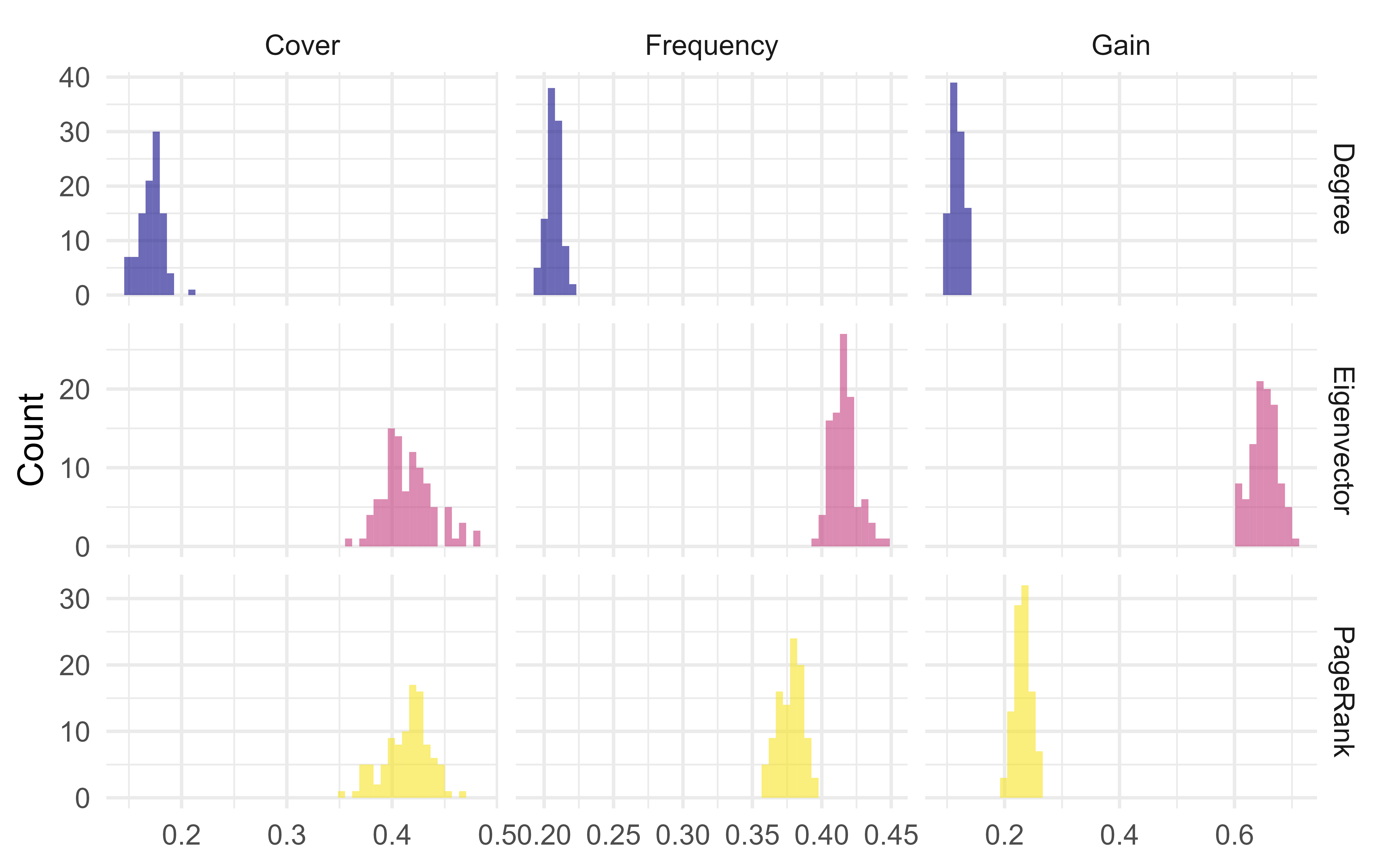}
    \caption{Distributions of variable importance as measured by Cover, Frequency, and Gain across 100 epidemic simulations for models including multi-layer centrality measures.}
    \label{fig:xgb-vi-mult}
\end{figure}

\begin{figure}[ht]
    \centering
    \includegraphics[scale=.9]{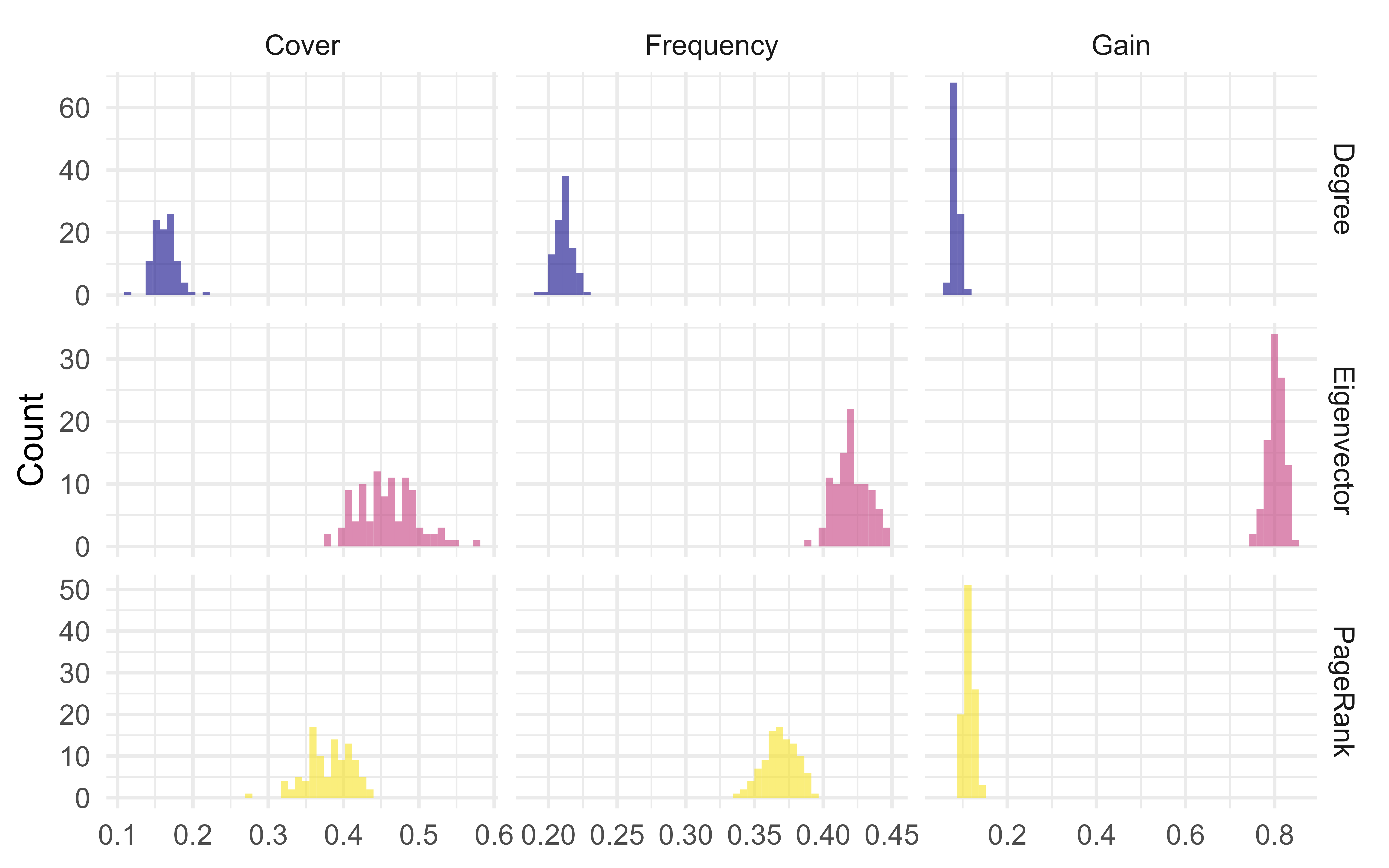}
    \caption{Distributions of variable importance as measured by Cover, Frequency, and Gain across 100 epidemic simulations for models including single-layer centrality measures.}
    \label{fig:xgb-vi-single}
\end{figure}

To assess how the individual centrality measures contributed to model performance, Figures \ref{fig:xgb-vi-mult} and \ref{fig:xgb-vi-single} depict distributions of three measures of variable importance metrics (Cover, Frequency and Gain, see Methods) across simulations.

In line with the results of the previous analyses, Eigenvector centrality contributed most to the performance of the XGBoost models, whereas PageRank, and most notably Degree centrality, did not add much to the predictions. This finding applies to both multi-layer and single-layer measures.

Results of the same prediction task using linear regression are provided in Supplementary Figure S5. Polynomial combinations of all centrality measures were required to achieve a maximum performance of $R^2=0.10, RMSE=2.76$, which lies notably below the results obtained from the XGBoost approach.

\subsubsection{XGBoost models with real infection data and additional predictors}

Using positive PCR-test results of our sample observations, we applied the XGBoost algorithm again to inspect the extent to which the centrality measures can be used to predict infection timing in actual COVID-19 infection data. However, neither using the model parameters trained on the simulated data nor training another set of models on the PCR-test data achieved any notable prediction performance for both the multi- and single-layer measures ($R^2=0.002, RMSE=14.86$). This indicates that administrative network data is  not sufficient to derive the actual infection process that relies on in-person contacts.

We thus explored how other personal characteristics could potentially improve predictions. We added age, the first two digits of the postcode area, and a variable indicating whether a person is of Dutch origin to the XGBoost models with the centrality measures. This lead to a small increase in model performance, to the same extent for both centrality types ($R^2=0.02, RMSE=14.67$). While this increase is still of no practical relevance, it should be noted that age was the most important predictor in these models according to a Gain value of 0.43. This relationship of age and time until infection also becomes apparent by comparing the infection process of different age groups as shown in Supplementary Figures S6 and S7.

\section{Conclusion}\label{sec:conclusion} 

In this study, we set out to investigate the predictive ability of centrality measures regarding the risk and timing of infections with a COVID-19-like disease. Drawing on large-scale registry data, we found that, among the investigated measures, Eigenvector centrality was best suited to predict risk and timing of individual infections. Especially relative infection risks could be identified well, even in the most parsimonious models. In practice, this would allow to identify individuals belonging to groups at high risk of infection solely based on their network position, and potentially enable targeted policies. Regarding the infection timing, predictive performance showed to be more limited, even when using a powerful machine learning algorithm like XGBoost. The estimated time of infection deviated on average around 2.3 weeks from the actual time point under the simulated epidemic scenario. Considering that the majority of simulated infections occurred within about 10 weeks (see Supplementary Figure S1) this only gives a rough indication of a person's infection timing.

Across all analyses, findings did not provide evidence for the multi-layer versions of centrality measures being better predictors of individual infections than their single-layer definitions. In fact, the opposite result emerged, with single-layer centrality measures being somewhat more powerful predictors. This raises the question in how far the more complex multi-layer representation of administrative social networks can contribute to understanding epidemic infection scenarios better.

Our analysis using positive PCR-tests indicates that registry data is not well suited to estimate infection risk at the individual level. While the discrepancy between real SARS-CoV-2 test data and the simulation could lie in the specification of the epidemic model, it likely resulted from the administrative data not reflecting social contacts accurately enough. Future research should investigate this issue more closely before conducting similar analyses, possibly combining the registry data with data based on contact tracing. Given that a more representative contact network can be established, centrality measures present an opportunity to augment models of infection prediction. Combining them with other individual characteristics related to infection risk, e.g., age or occupation, might result in a more powerful tool to inform targeted government interventions or provide guidelines for individuals.

While we only worked with a subset of the population data, the network characteristics such as the Degree distribution resembled closely those of the complete population network as discussed in \cite{vanderlaan2023}. Results from this study can thus likely be transferred to a population-based analysis. 

Our analysis has a number of limitations that open up additional avenues for future research. Given the size of the data---ca. 1.6M nodes and 61M ties in total---and the increased computational power needed for working with multi-layer networks, we only were able to use centrality measures based on random walks or a node's direct neighborhood. However, measures based on shortest paths, like Betweenness or Closeness, could offer new insights. Calculating these measures would require the use of approximate methods.

Another immediate opportunity would be to test the presented measures in varying epidemic scenarios. This could be done by either varying the parameters of the here used SIR model, as well as considering other frameworks of epidemic disease modeling, e.g., allowing for additional infection states. Similarly, the same models could be applied to different regions or countries for which comparable data is available. An extension to a time-dependent approach is another possibility. This would allow to account for changing spreading dynamics caused by interventions, vaccination, or other time-dependent factors.

Apart from other applications and modeling approaches, more fundamental work is needed to understand why certain measures perform better or worse than others. For the simulated epidemics, the higher performance of Eigenvector centrality can partly be attributed to the explicit inclusion of an individual's direct neighborhood in generating new infections. Still, it would be valuable to derive exactly how the centrality measures relate to epidemic outcomes.

\bibliography{references}

\newpage

\section*{Acknowledgements}

The project received funding from the Dutch Ministry of Health, Welfare and Sport, and the Open Data Infrastructure for Social Science and Economic Innovations (ODISSEI). We thank Tom Emery, Vincent Buskens, and Albert-Jan van Hoek for their continued support as well as valuable discussions and comments.

\section*{Author contributions}

Conceptualization: CH, JG, and AB. Data curation and analysis: CH. Visualization: CH. Writing---original draft: CH. Writing---review and editing: CH, JG, and AB.

\section*{Competing interests}

The authors declare no competing interests.

\section*{Data and code availability}

Results are based on calculations by the authors using non-public micro-data from Statistics Netherlands. Under certain conditions, this micro-data is accessible for statistical and scientific research. For further information contact microdata@cbs.nl. Please see also \cite{bokanyi2023} for more explanation on the data, as well as \cite{vanderlaan2023} for access and usage regulations. The code documenting the data processing and analyses of this study is available at \url{https://github.com/christine-hvw/thesis_disclosed/tree/publication}.

\subsection*{CBS datasets used in this study}

\begin{itemize}
    \item PersNw2018\_v1.0\_links\_familie
    \item PersNw2018\_v1.0\_links\_huishoude
    \item PersNw2018\_v1.0\_links\_werk
    \item PersNw2018\_v1.0\_linktype\_labels
    \item BRINADRESSEN2020V1
    \item INSCHRWPOTAB2020V2
    \item CoronIT\_GGD\_testdata\_20210921
\end{itemize}

\newpage

\section*{Supplementary Information}

\appendix

\setcounter{table}{0}
\renewcommand{\thetable}{S\arabic{table}}
\setcounter{figure}{0}
\renewcommand{\thefigure}{S\arabic{figure}}

\section{Descriptive statistics of epidemic simulations}

In total, 100 simulations of a SIR epidemic were performed on the network (see main text Section 4.1). Table \ref{tab:descriptives_sim} below lists descriptive statistics of the simulations.

\begin{table}[ht]
\centering
\begin{threeparttable}
\caption{Descriptive statistics of simulated epidemic SIR process on the network.}
\addtolength{\tabcolsep}{-1pt}
\begin{tabular}{@{}llllll@{}}
\toprule
                                & Min.   & Mean  & Median& Max     & SD    \\ \midrule
Duration (weeks)                & 33     & 37.20 & 37    & 43      & 2.12   \\
Total nodes infected (\%)       & 68.86  & 69.45 & 69.46 & 70.10   & 0.26   \\
Avg. time to infection (weeks)  & 12.26  & 13.93 & 13.85 & 16.32   & 0.74   \\ 
Infections at peak (\%)         & 12.24  & 12.96 & 12.97 & 13.87   & 0.34   \\ 
Time of infection peak (week)   & 13     & 14.48 & 14    & 17      & 0.85   \\ 
Infections per node (\%)        & 0.00   & 69.45 & 82.00 & 100.00  & 33.15  \\ \bottomrule
\end{tabular}
\label{tab:descriptives_sim}
\end{threeparttable}
\end{table}

\begin{figure}[ht]
    \centering
    \includegraphics[width=\textwidth]{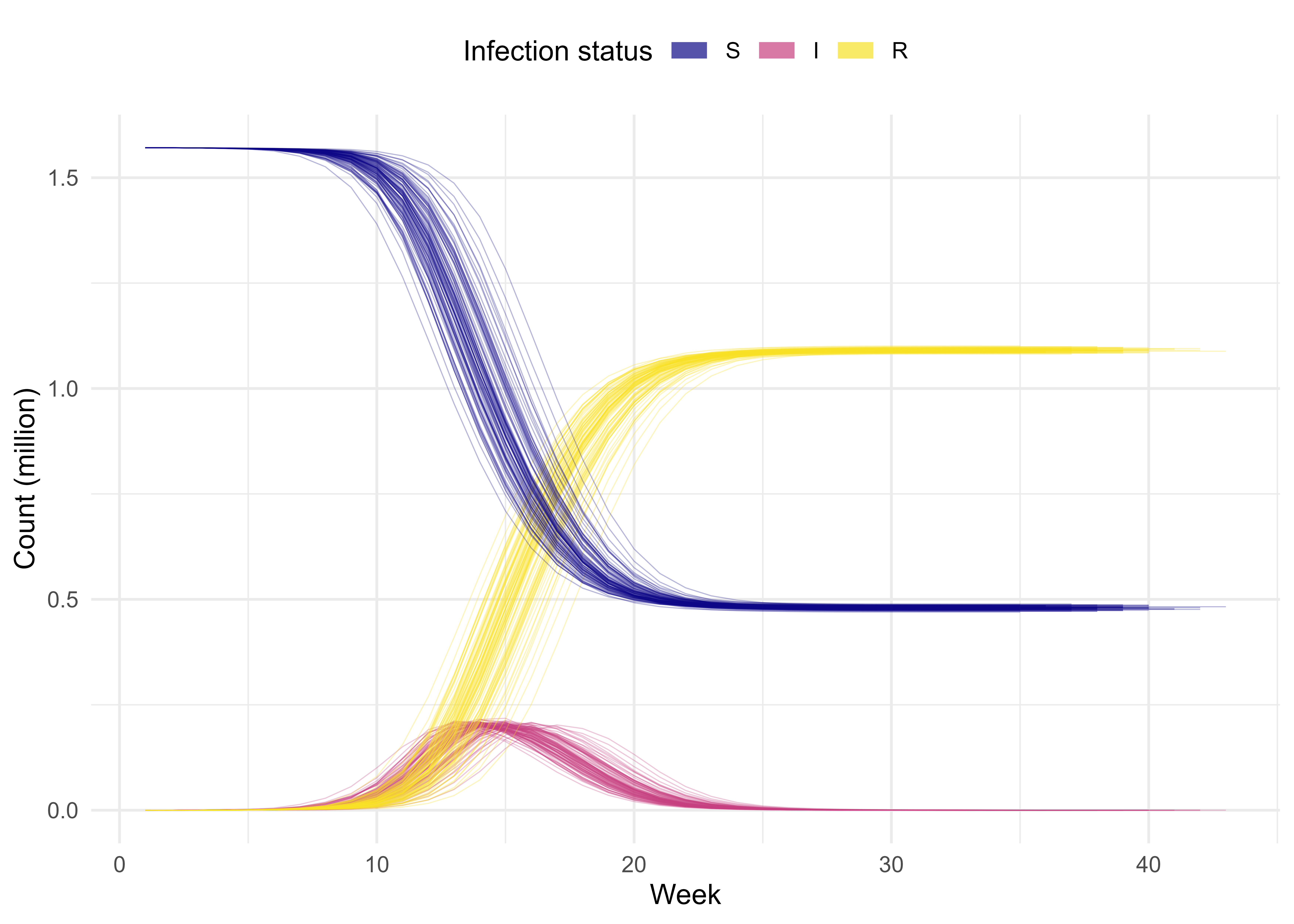}
    \caption{Counts of infection status groups for 100 simulations of a SIR model.}
    \label{fig:epi-curves}
\end{figure}

The minimum of 0 infections per node means that some nodes were not infected in any of the simulations. However, this was the case only for 0.28\% of all nodes. Figure \ref{fig:epi-curves} summarizes the epidemic process for all simulations by presenting the number of nodes belonging to the different infection status groups (susceptible, infected, removed) over time.

\clearpage

\section{Distributions of centrality measures}

\begin{figure}[ht]
    \centering
    \includegraphics[scale=0.9]{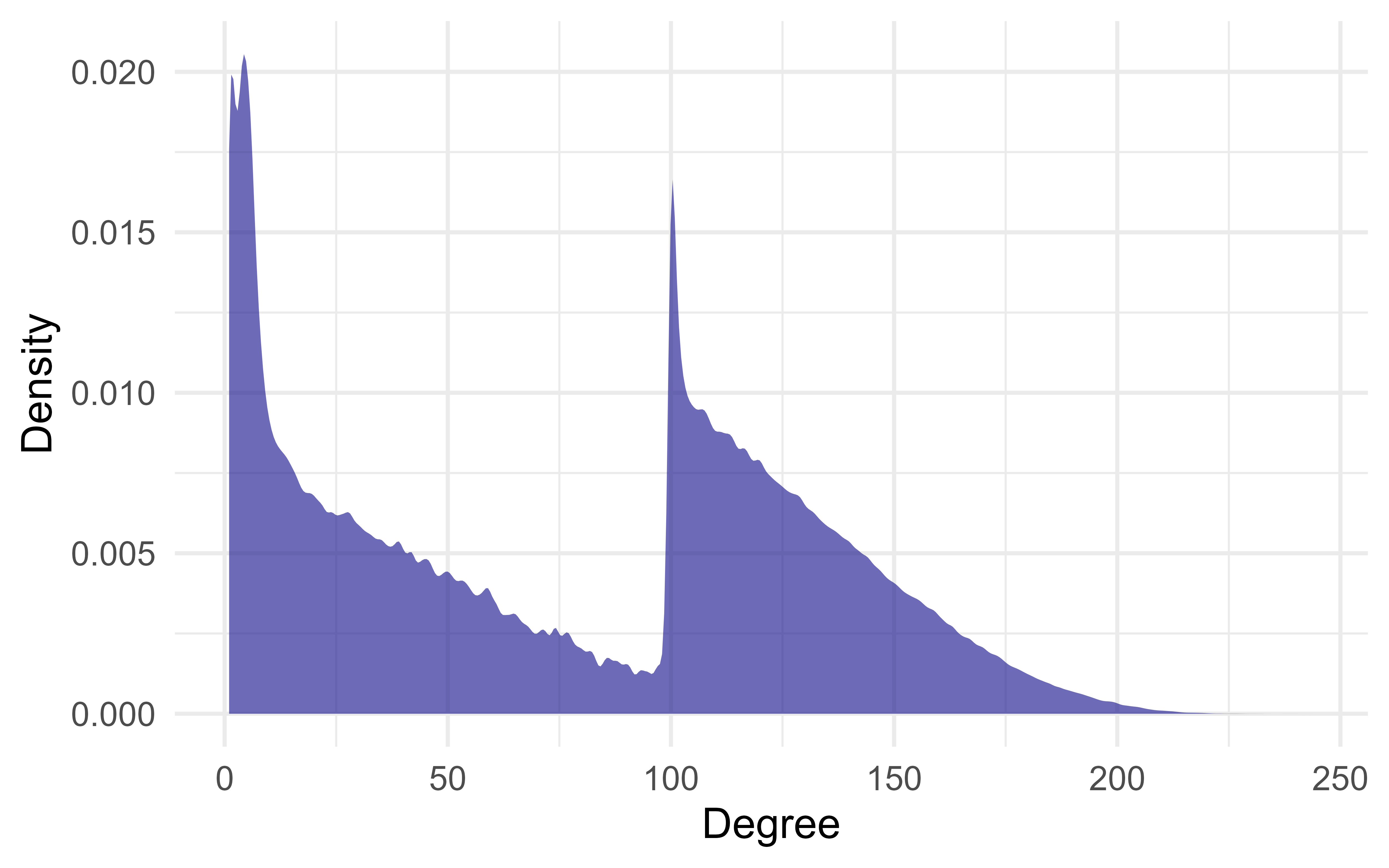}
    \caption{Density plot of Degree distribution.}
    \label{fig:dist-deg}
\end{figure}

\begin{figure}[ht]
    \centering
    \includegraphics[scale=1]{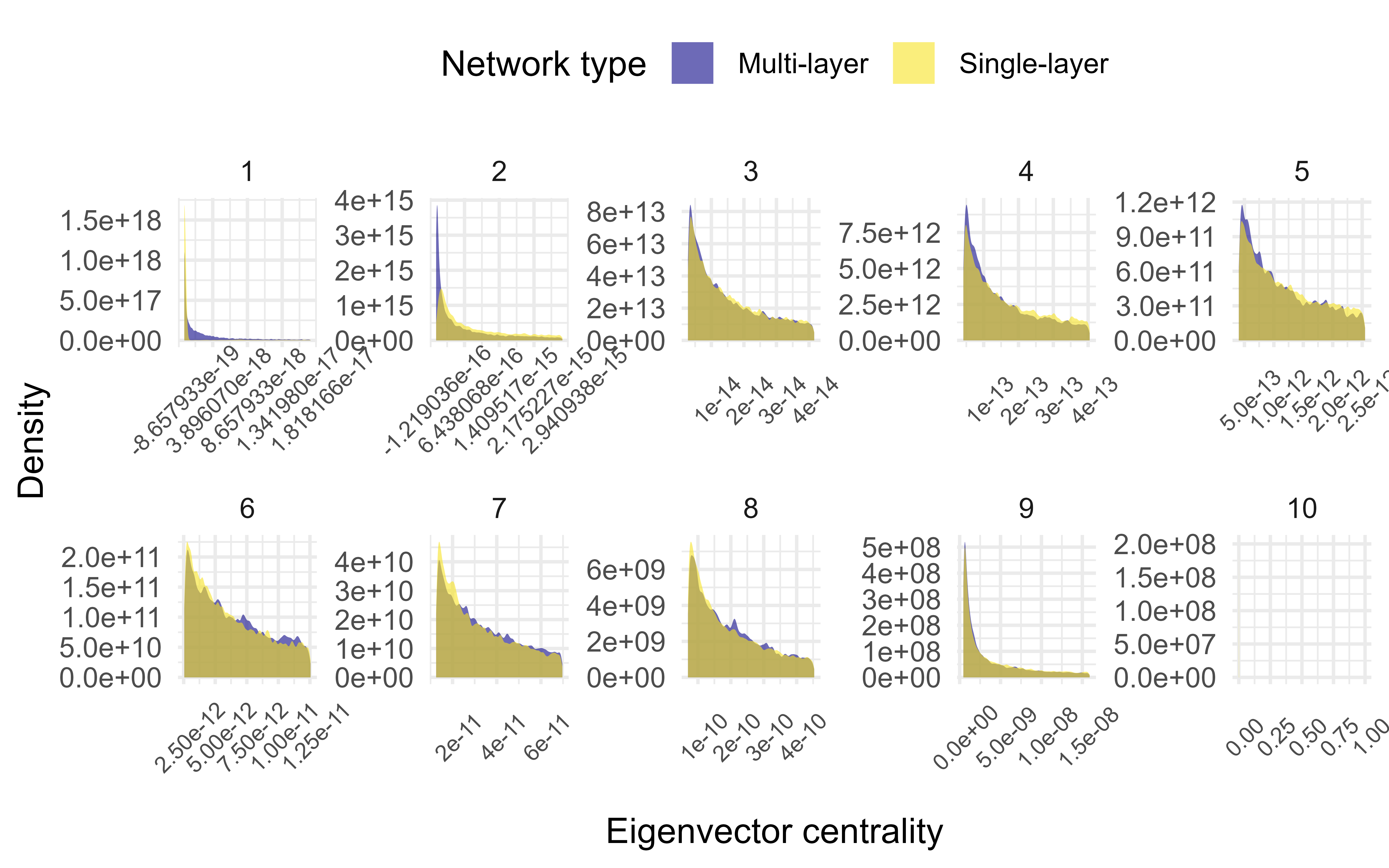}
    \caption{Density plot of Eigenvector centrality distribution by centrality decile and type of network structure.}
    \label{fig:dist-eig}
\end{figure}

\begin{figure}[ht]
    \centering
    \includegraphics[scale=1]{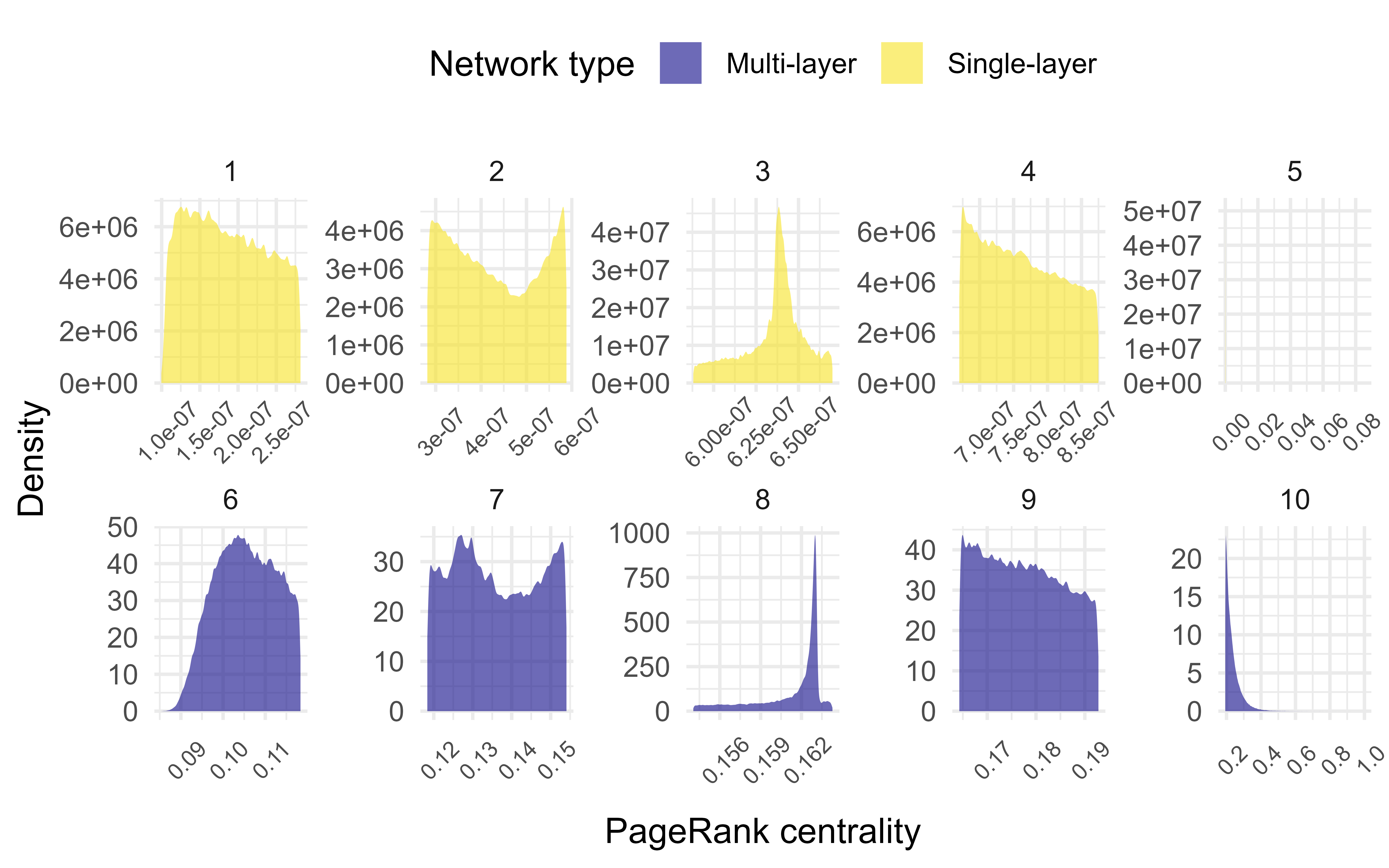}
    \caption{Density plot of PageRank centrality distribution by centrality decile and type of network structure.}
    \label{fig:dist-pr}
\end{figure}

\clearpage

\section{Hyperparameter settings of XGBoost models}

\begin{table}[ht]
\centering
\begin{threeparttable}
\caption{Hyperparameter settings of XGBoost models.}
\begin{tabular}{@{}ll@{}}
\toprule
Hyperparameter                   & Setting        \\ \midrule
Trees                            & 500            \\
Min. node size                   & 20             \\
Max. tree depth                  & 4              \\
Learning rate                    & 0.1            \\
Minimum loss reduction           & 0              \\
No. randomly selected predictors & No. predictors \\
Proportion of sampled obs.       & 0.5            \\ \bottomrule
\end{tabular}
\label{tab:hyperparams}
\end{threeparttable}
\end{table}

\section{Additional statistics of Cox proportional hazard models}

\begin{table}[H]
\centering
\begin{threeparttable}
\caption{Mean and confidence intervals of Concordance indices of Cox proportional hazard models across simulations.}
\addtolength{\tabcolsep}{-1pt}
\begin{tabular}{lllll}
\hline
                           & \multicolumn{2}{l}{Multi-layer}  & \multicolumn{2}{l}{Single-layer} \\ \hline
\multicolumn{1}{l|}{Model} & Mean          & 95\% CI          & Mean          & 95\% CI          \\ \hline
\multicolumn{1}{l|}{1}     & 0.68          & [0.682, 0.683]   & 0.68          & [0.682, 0.683]   \\
\multicolumn{1}{l|}{2}     & 0.68          & [0.681, 0.682]   & 0.68          & [0.681, 0.682]   \\
\multicolumn{1}{l|}{3}     & 0.68          & [0.682, 0.683]   & 0.68          & [0.682, 0.683]   \\
\multicolumn{1}{l|}{4}     & 0.71          & [0.713, 0.714]   & 0.77          & [0.768, 0.770]   \\
\multicolumn{1}{l|}{5}     & 0.71          & [0.713, 0.714]   & 0.77          & [0.768, 0.770]   \\
\multicolumn{1}{l|}{6}     & 0.71          & [0.713, 0.714]   & 0.77          & [0.768, 0.770]   \\
\multicolumn{1}{l|}{7}     & 0.62          & [0.622, 0.622]   & 0.65          & [0.650, 0.651]   \\
\multicolumn{1}{l|}{8}     & 0.63          & [0.632, 0.633]   & 0.66          & [0.657, 0.657]   \\
\multicolumn{1}{l|}{9}     & 0.64          & [0.637, 0.638]   & 0.66          & [0.660, 0.660]   \\
\multicolumn{1}{l|}{10}    & 0.68          & [0.683, 0.684]   & 0.68          & [0.682, 0.683]   \\
\multicolumn{1}{l|}{11}    & 0.69          & [0.691, 0.692]   & 0.69          & [0.686, 0.686]   \\
\multicolumn{1}{l|}{12}    & 0.62          & [0.623, 0.624]   & 0.65          & [0.650, 0.651]   \\
\multicolumn{1}{l|}{13}    & 0.69          & [0.691, 0.692]   & 0.69          & [0.686, 0.686]   \\
\multicolumn{1}{l|}{14}    & 0.68          & [0.685, 0.685]   & 0.68          & [0.682, 0.683]   \\
\multicolumn{1}{l|}{15}    & 0.70          & [0.704, 0.705]   & 0.70          & [0.700, 0.701]   \\
\multicolumn{1}{l|}{16}    & 0.63          & [0.635, 0.635]   & 0.66          & [0.657, 0.657]   \\
\multicolumn{1}{l|}{17}    & 0.70          & [0.704, 0.705]   & 0.70          & [0.700, 0.701]   \\
\multicolumn{1}{l|}{18}    & 0.68          & [0.685, 0.685]   & 0.68          & [0.683, 0.684]   \\
\multicolumn{1}{l|}{19}    & 0.71          & [0.706, 0.706]   & 0.70          & [0.699, 0.700]   \\
\multicolumn{1}{l|}{20}    & 0.64          & [0.640, 0.641]   & 0.66          & [0.660, 0.660]   \\
\multicolumn{1}{l|}{21}    & 0.71          & [0.706, 0.707]   & 0.70          & [0.699, 0.700]   \\
\multicolumn{1}{l|}{22}    & 0.68          & [0.683, 0.684]   & 0.68          & [0.683, 0.684]   \\
\multicolumn{1}{l|}{23}    & 0.71          & [0.709, 0.710]   & 0.71          & [0.705, 0.706]   \\
\multicolumn{1}{l|}{24}    & 0.62          & [0.624, 0.624]   & 0.65          & [0.650, 0.651]   \\
\multicolumn{1}{l|}{25}    & 0.71          & [0.708, 0.709]   & 0.71          & [0.705, 0.706]   \\ \hline
\end{tabular}
\label{tab:cox-results-cis}
    \begin{tablenotes}
      \small
      \item \textit{Note:} Full model specifications are displayed in main text Figure 2.
    \end{tablenotes}
  \end{threeparttable}
\end{table}

\clearpage

\section{Linear regression models}

\begin{figure}[ht]
    \centering
    \includegraphics[scale=.9, trim=.8cm 3cm .05cm .1cm, clip]{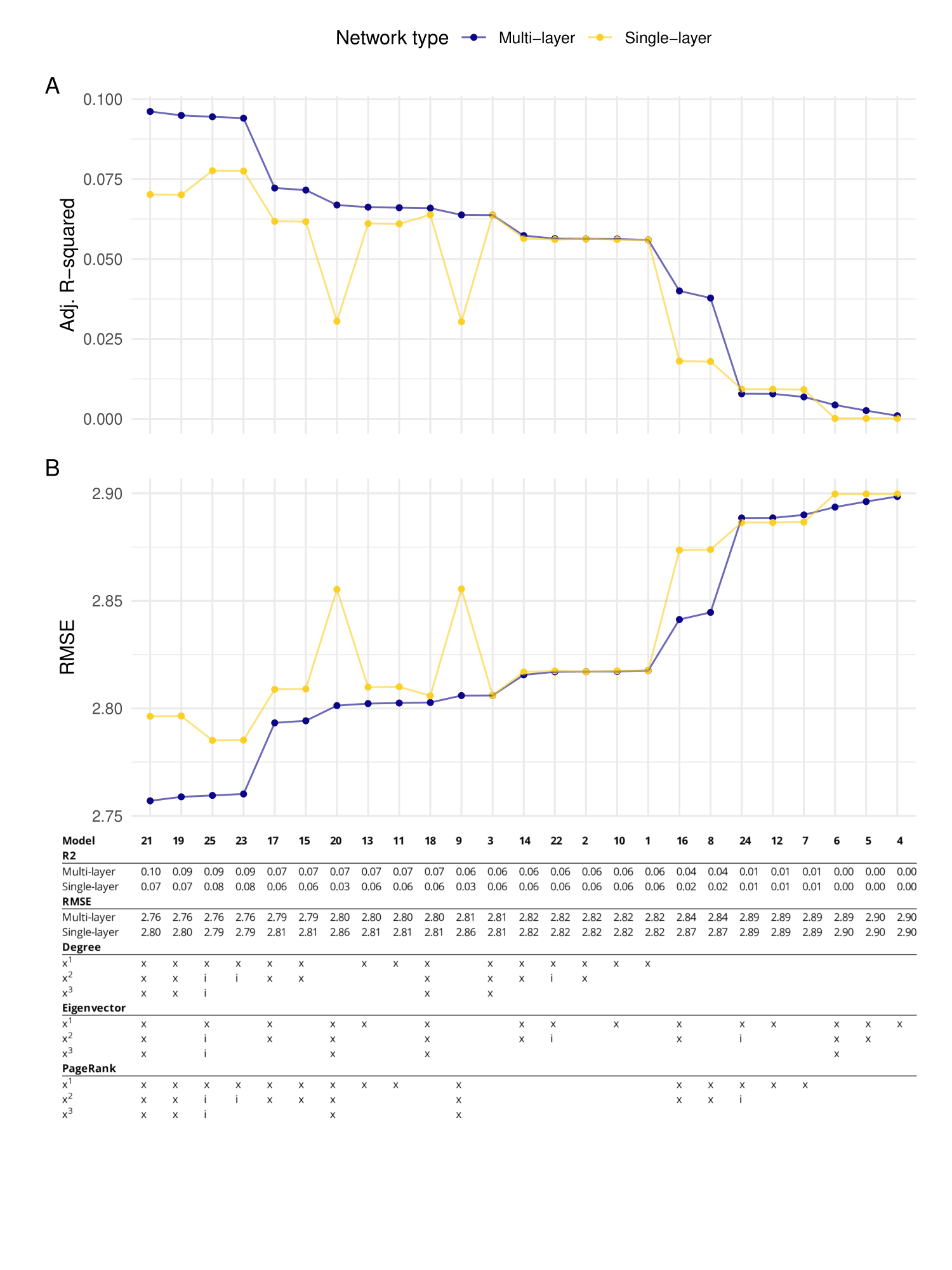}
    \caption{Performance measures of linear regression models by type and order of included centrality measure. \textbf{A} Adjusted R-square values. \textbf{B} Root-mean-square error values. The table below applies to both panels A and B. An \textit{x} denotes terms composed of a single variable, \textit{i} denotes two- and three-way interactions between multiple variables. Since the estimates' confidence intervals resulting from the simulations were very narrow, they were omitted from this figure.}
    \label{fig:reg-results}
\end{figure}

\clearpage

\section{Descriptive statistics of observations in PCR-test data analyses}

\begin{table}[ht]
\centering
\begin{threeparttable}
\caption{Descriptive statistics of sample observations with positive PCR-test.}
\addtolength{\tabcolsep}{-1pt}
\begin{tabular}{@{}llllll@{}}
\toprule
                                & Pctl. 25 & Mean  & Median& Pctl. 75& SD    \\ \midrule
Time until infection (weeks)    & 24     & 36.84   & 35    & 48      & 14.85   \\
Year of birth                   & 1970   & 1981.53 & 1982  & 1993    & 13.90   \\
Avg. time to infection (weeks)  & 12.26  & 13.93   & 13.85 & 16.32   & 0.74   \\ 
Dutch origin                    &        & 0.60    &       &         &        \\ 
Degree centrality               & 22     & 78.70   & 92    & 124     & 55.72   \\
Eigenvector centrality (multi)  &        & 0.0005  &       &         & 0.01   \\
Eigenvector centrality (single) &        & 0.0001  &       &         & 0.01   \\
PageRank centrality (multi)     &        & 0.16    &       &         & 0.05   \\
PageRank centrality (single)    &        & 6.38e-7 &       &         & 3.69e-07 \\ \bottomrule
\end{tabular}
\begin{tablenotes}
  \small
  \item \textit{Note:} Instead of minima and maxima, the lower and upper 25 percentiles are given as to not disclose information at the individual level. Cells are empty for values applying to fewer than ten individuals, or where no other statistics are applicable (indicator variable Dutch origin).
\end{tablenotes}
\label{tab:descriptives_pcr}
\end{threeparttable}
\end{table}

\begin{figure}[ht]
    \centering
    \includegraphics[scale=.9]{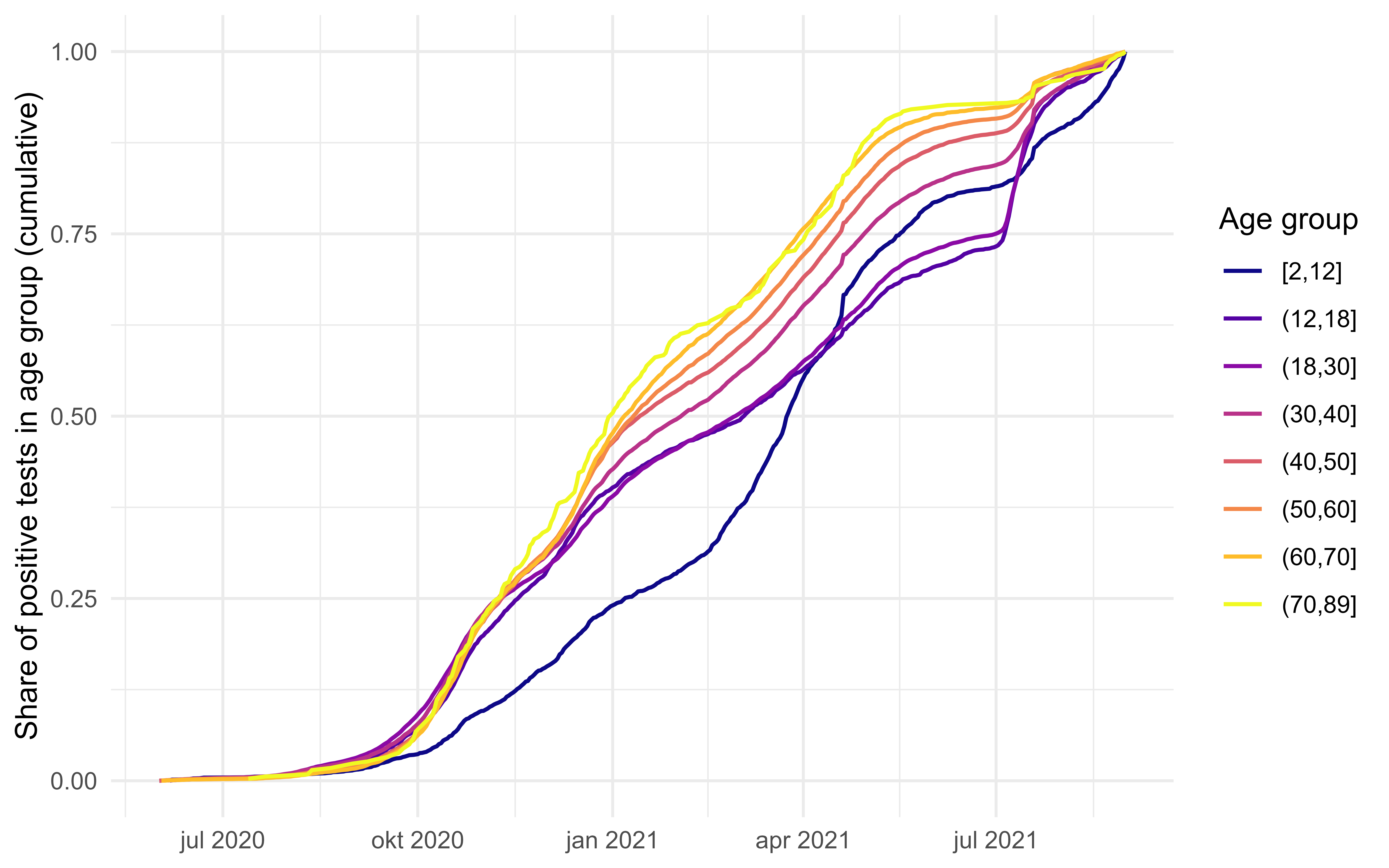}
    \caption{Cumulative share of positive PCR-test results within age groups over time.}
    \label{fig:pcr-byage}
\end{figure}

\begin{figure}[ht]
    \centering
    \includegraphics[scale=.9]{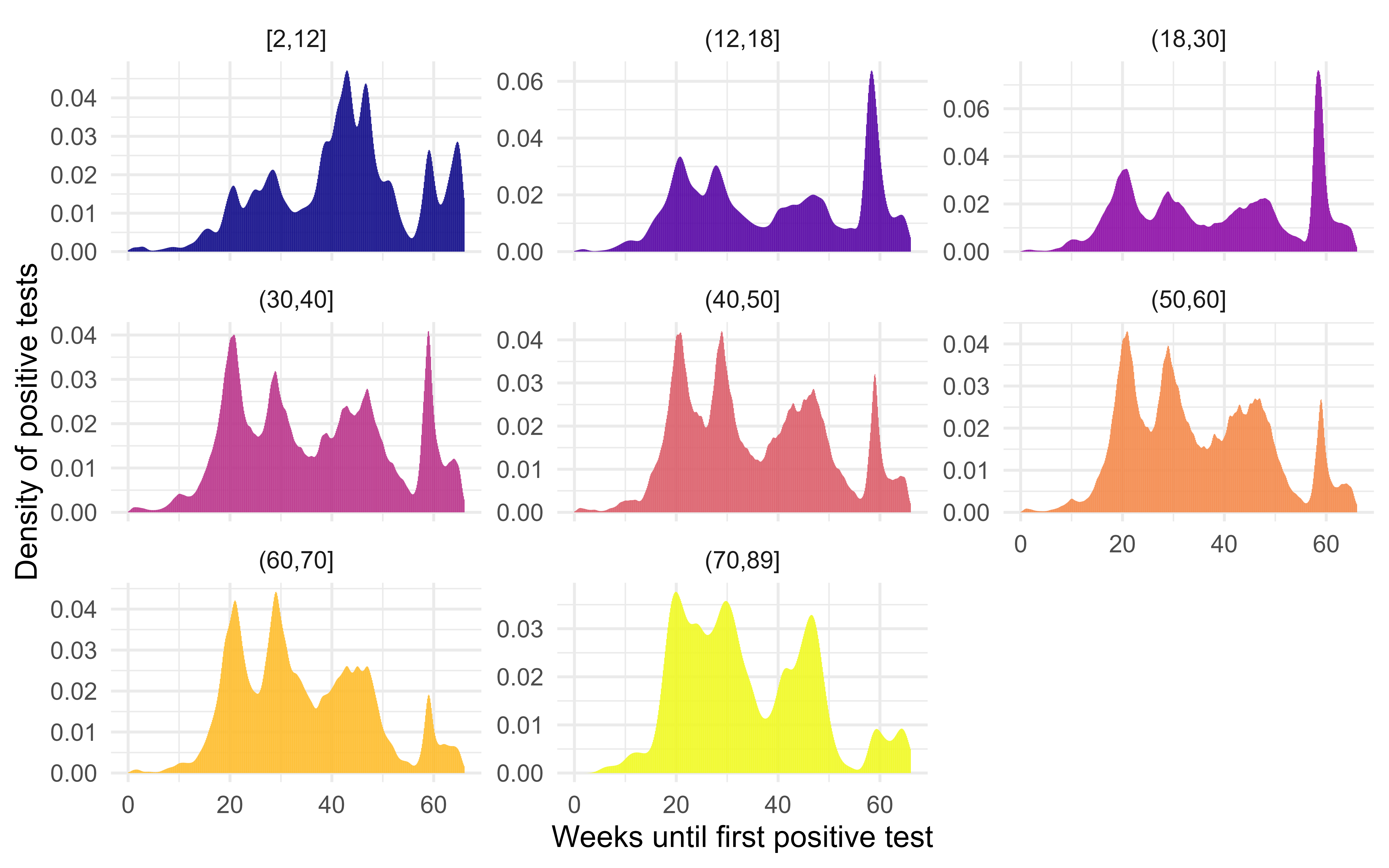}
    \caption{Distributions of time until infection (weeks) as measured by first positive test result after June 1, 2020, by age group.}
    \label{fig:tti_byage}
\end{figure}

\end{document}